\documentclass[10pt,conference]{style/IEEEtran}

\usepackage{cite}
\usepackage{amsmath,amssymb,amsfonts}
\usepackage{graphicx}
\usepackage{booktabs}
\usepackage{multirow}
\usepackage{tabularx}
\usepackage{makecell}
\usepackage{array}
\usepackage{enumitem}
\usepackage{textcomp}
\usepackage{float}
\usepackage[table]{xcolor}
\usepackage{url}

\graphicspath{{figures/}}

\def\BibTeX{{\rm B\kern-.05em{\sc i\kern-.025em b}\kern-.08em
    T\kern-.1667em\lower.7ex\hbox{E}\kern-.125emX}}

\begin{document}

\title{His2Trans: A Knowledge-Guided Agentic Framework for Project-Level C-to-Rust Migration}

\author{
\IEEEauthorblockN{
Shengbo Wang,
Mingwei Liu\IEEEauthorrefmark{1},
Guangsheng Ou,
Yuwen Chen,
Zike Li,
Yanlin Wang,
and Zibin Zheng
}

\IEEEauthorblockA{
Sun Yat-sen University\\
Guangzhou, China\\
\small
\begin{tabular}{c}
puppytagge@gmail.com,
liumw26@mail.sysu.edu.cn,
ougsh3@mail2.sysu.edu.cn,\\
chenyw95@mail2.sysu.edu.cn,
lizk8@mail2.sysu.edu.cn,
wangylin36@mail.sysu.edu.cn,
zhzibin@mail.sysu.edu.cn
\end{tabular}
}

\thanks{\IEEEauthorrefmark{1}Corresponding author.}
}

\maketitle

\begin{abstract}
C remains a major implementation language for operating systems, embedded platforms, and infrastructure software, but manual memory management continues to create security and maintenance costs. Rust is a practical migration target because it retains low-level control while enforcing stronger memory-safety checks. At project scale, especially under gradual C/Rust coexistence, migration is not a sequence of syntax-preserving function rewrites. A translator must preserve project interfaces, observable behavior, system interaction protocols, and low-level interoperability boundaries while staying consistent with migration choices already made in the codebase. 

We introduce His2Trans, a knowledge-guided agentic framework for project-level C-to-Rust migration. His2Trans reuses interface-level and fragment-level knowledge mined from historical C/Rust migrations to guide new translations toward Rust interfaces, wrapper choices, and local idioms already accepted in the evolving project. It then refines the assembled crate with project-level agentic feedback. On ten OpenHarmony modules, His2Trans reaches a 100.00\% incremental compilation pass rate, a 94.92\% Test Pass Rate, and a 16.35\% Unsafe Ratio. On eight open-source C projects, it reaches 100.00\% for both incremental compilation and Test Pass Rate, reducing Unsafe Ratio from 42.88\% under C2Rust to 8.59\%. These results support knowledge-guided migration and project-level agentic refinement as practical mechanisms for preserving observable behavior while reducing the unsafe burden of rule-based transpilation.

\end{abstract}

\begin{IEEEkeywords}
C-to-Rust, large language models, historical knowledge reuse, project-level migration, unsafe reduction
\end{IEEEkeywords}

\section{Introduction}
\label{sec:intro}

C remains central to operating systems, embedded runtimes, device drivers, protocol stacks, and infrastructure software. Its control over memory layout, pointer arithmetic, and hardware resources also leaves developers responsible for avoiding buffer overflows, use-after-free errors, null dereferences, and data races. Industry reports have repeatedly attributed a large fraction of severe vulnerabilities to memory-safety defects~\cite{Miller2019Trends,ChromiumMemorySafety}. Rust has therefore become an attractive migration target for systems software: it keeps low-level control while moving many memory-safety checks into the compiler~\cite{10.1145/3418295,10.1145/3158154}.

Automated C-to-Rust migration has followed two paths. Rule-based tools such as C2Rust~\cite{Galois2018C2Rust} can translate substantial C code, but they preserve much of the C execution structure through raw pointers and extensive \texttt{unsafe} code~\cite{10.1145/3485498}. Large-language-model-based methods instead use model-generated Rust, dependency graphs, compiler/test-guided repair, or static-analysis feedback to improve correctness and idiomaticity~\cite{articlerustflow,nitin2025spectraenhancingcodetranslation,10.1145/3695988,wang2026evoc2rustskeletonguidedframeworkprojectlevel,yuan2025projectlevelctorusttranslationsynergistic,10.1007/978-3-032-00828-2_16}. These systems make clear that project-level migration cannot be reduced to isolated function translation. 

\textbf{Yet compilable Rust is still not enough in gradual C/Rust migration.} Such migrations usually unfold inside codebases where C and Rust coexist for long periods: some subsystems have already migrated, some entry points are still called from C, and reused Rust wrappers or externally observable protocols have become part of the project surface. A new Rust artifact must therefore fit these existing migration choices, especially public interfaces, entries called across the C/Rust boundary, and externally visible protocols. Examples include service-manager interfaces, inter-process communication wire formats, and driver callbacks, among others~\cite{AndroidMemorySafetyAOSP,ChromiumOSDevelopmentBasics,LinuxKernelRustQuickStart,FirefoxRustCppInterop}. Such interfaces and protocols become sources of evidence once previous migrations have handled them. The resulting C/Rust pairs record reusable decisions: which Rust-side wrapper replaced a C interface, how error codes were propagated, which field order an inter-process communication buffer preserved, and other project-specific choices. Later translations should reuse them to remain consistent with the evolving C/Rust codebase.

Without migration-specific evidence and an explicit boundary, a general-purpose coding agent may produce a Rust crate that is locally plausible but incompatible with the original project surface. This gap appears in our OpenHarmony evaluation: Claude Code-only reaches 100.00\% incremental compilation, yet its test pass rate drops to 80.08\%. The failed cases involve system-specific contracts such as service-manager linkage, hardware-configuration parsing, device callbacks, and other protocols not enforced by compilation.

To address this gap, we introduce His2Trans, a knowledge-guided agentic framework for project-level C-to-Rust migration. His2Trans fixes the migration boundary and target crate structure, then uses historical migration knowledge to guide function translation. Interface-level knowledge helps the model reuse project-compatible Rust interfaces, and fragment-level knowledge captures recurring implementation idioms. A final project-level refinement stage uses compilation, semantic, and unsafe feedback to improve the assembled artifact. The goal is to produce Rust migration artifacts that compile, preserve observable behavior, remain consistent with prior migration choices, keep low-level interoperability boundaries explicit, and reduce measured unsafe code within a reproducible evaluation scope.

Table~\ref{tab:comparison} contrasts these paradigms and highlights the gap His2Trans targets: combining explicit migration boundaries, historical C/Rust evidence, and project-level refinement in one agentic workflow.

\begin{table*}[t]
\caption{Positioning of His2Trans among C-to-Rust migration paradigms.}
\label{tab:comparison}
\scriptsize
\setlength{\tabcolsep}{2.5pt}
\renewcommand{\arraystretch}{1.2}
\renewcommand\tabularxcolumn[1]{p{#1}}
\newcolumntype{Y}{>{\raggedright\arraybackslash}X}
\begin{tabularx}{\textwidth}{@{}p{0.15\textwidth}p{0.24\textwidth}p{0.14\textwidth}Y Y@{}}
\toprule
\textbf{Paradigm} & \textbf{Representative methods} & \textbf{Primary target} & \textbf{Main idea} & \textbf{Relation to His2Trans} \\
\midrule
Rule-based transpilation & C2Rust~\cite{Galois2018C2Rust}, ownership/safety rewrites~\cite{10.1145/3485498,10.1007/978-3-031-37709-9_22,emre2023aliasing,ling2022rust} & Files / projects & Preserve C structure while lowering syntax and memory operations & Predictable coverage, but often C-style unsafe; His2Trans adds historical knowledge and refinement \\
\addlinespace[0.3em]
Function-level LLM-based translation & SafeTrans~\cite{farrukh2025safetransllmassistedtranspilationc}, SPECTRA~\cite{nitin2025spectraenhancingcodetranslation}, VERT~\cite{yang2024vertverifiedequivalentrust}, RustFlow~\cite{articlerustflow} & Functions / snippets & Generate isolated units with compile/test/specification feedback & Flexible at unit level; His2Trans targets project boundaries and cross-module protocols \\
\addlinespace[0.3em]
Structured project-level LLM-based migration & EvoC2Rust~\cite{wang2026evoc2rustskeletonguidedframeworkprojectlevel}, PTRMAPPER~\cite{yuan2025projectlevelctorusttranslationsynergistic}, RustMap~\cite{10.1007/978-3-032-00828-2_16}, LLMigrate~\cite{liu2025llmigratetransforminglazylarge}, Tymcrat~\cite{Tymcrat} & Projects / modules & Organize translation with skeletons, graphs, or type migration & Provides project context; His2Trans adds retrieved interface/fragment knowledge and refinement \\
\addlinespace[0.3em]
Agentic C-to-Rust migration & ORBIT~\cite{farrukh2026orbit}, RustPrint~\cite{leanh2026rustprint}, ENCRUST~\cite{sim2026encrust}, ACToR~\cite{li2025actor}, LAC2R~\cite{sim2025lac2r}, Rustify*~\cite{wang2025rustifywithdrawn} & Programs / repositories & Coordinate agents, scaffolds, validation, or search & Closest workflow line; His2Trans grounds decisions in historical C/Rust pairs \\
\addlinespace[0.3em]
\rowcolor{gray!12}
Knowledge-guided agentic migration & His2Trans & Project-level migration, especially build-complex and partially migrated C/Rust ecosystems & Retrieve interface/fragment knowledge, translate in a fixed scaffold, then refine & Targets testable artifacts with explicit migration boundaries and reduced unsafe usage \\
\bottomrule
\end{tabularx}
\vspace{0.25em}
\end{table*}

We evaluate His2Trans on two datasets: ten OpenHarmony modules from a build-complex, partially migrated systems ecosystem, and eight open-source C projects used to assess transfer beyond that ecosystem. On the OpenHarmony module dataset, His2Trans achieves 100.00\% incremental compilation, a 94.92\% test pass rate, and a 16.35\% Unsafe Ratio, outperforming the evaluated baselines in the combination of tested behavior and unsafe-code reduction. On the open-source project dataset, His2Trans reaches 100.00\% incremental compilation and 100.00\% test pass rate, while reducing Unsafe Ratio to 8.59\%, compared with 42.88\% for C2Rust.

This paper makes the following contributions:

\begin{enumerate}[label=(\arabic*),leftmargin=*]
\item We formulate project-level C-to-Rust migration in gradual mixed-language systems as a boundary-constrained, knowledge-guided agentic process, and implement it in His2Trans through scaffold construction, historically guided function translation, and project-level refinement.
\item We show how historical C/Rust migration pairs can be mined as interface-level and fragment-level guidance so that new Rust artifacts reuse project-compatible wrappers, local idioms, and system-facing protocols instead of relying only on locally plausible generated code.
\item We provide evidence across ten OpenHarmony modules and eight open-source C projects that this design preserves tested behavior while reducing measured unsafe code under explicit migration boundaries.
\end{enumerate}

\section{Methodology}
\label{sec:methodology}

\begin{figure*}[t]
\centering
\includegraphics[width=\textwidth, trim={0 11.6cm 14.3cm 0}, clip]{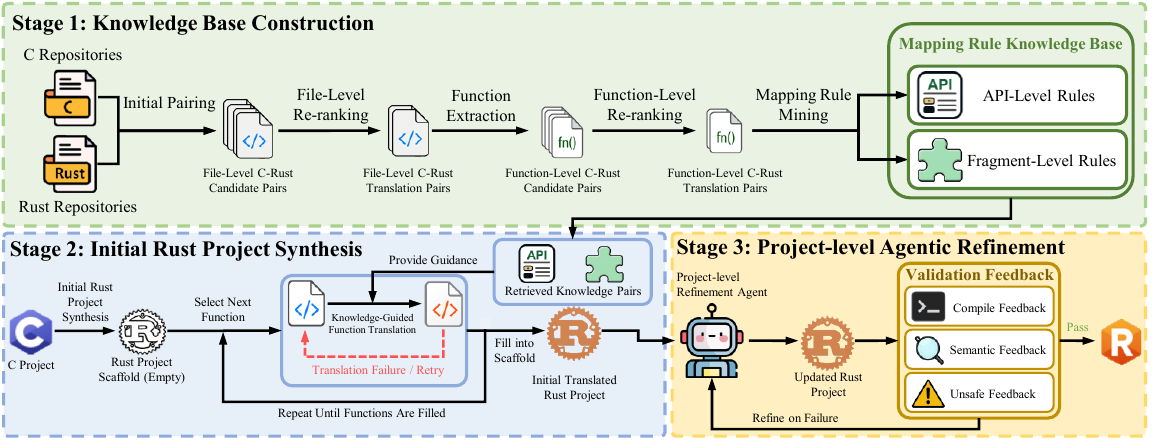}
\caption{Overview of His2Trans. Historical C/Rust examples are mined into interface-level and fragment-level knowledge; the target project is converted into a Rust scaffold; function bodies are translated with retrieved knowledge as guidance; and the assembled crate is refined using compilation, semantic, and unsafe feedback.}
\label{fig:framework}
\end{figure*}

His2Trans casts project-level C-to-Rust migration as constrained construction followed by refinement, rather than as free-form crate generation. Figure~\ref{fig:framework} summarizes the workflow. The first stage derives reusable migration knowledge from prior C/Rust examples. The second stage converts the target C project into a Rust scaffold and translates function bodies with guidance from the retrieved knowledge. The third stage refines the assembled crate with compilation, semantic, and unsafe feedback.

\subsection{Historical Knowledge Base Construction}
\label{sec:kb_construction}

The knowledge base stores reusable migration experience. His2Trans mines historical C/Rust repositories or co-evolved modules, aligns likely migration examples, and summarizes them into compact guidance for retrieval during function translation.

\begin{figure*}[t]
\centering
\includegraphics[width=\linewidth, trim={0 9.5cm 2.1cm 0}, clip]{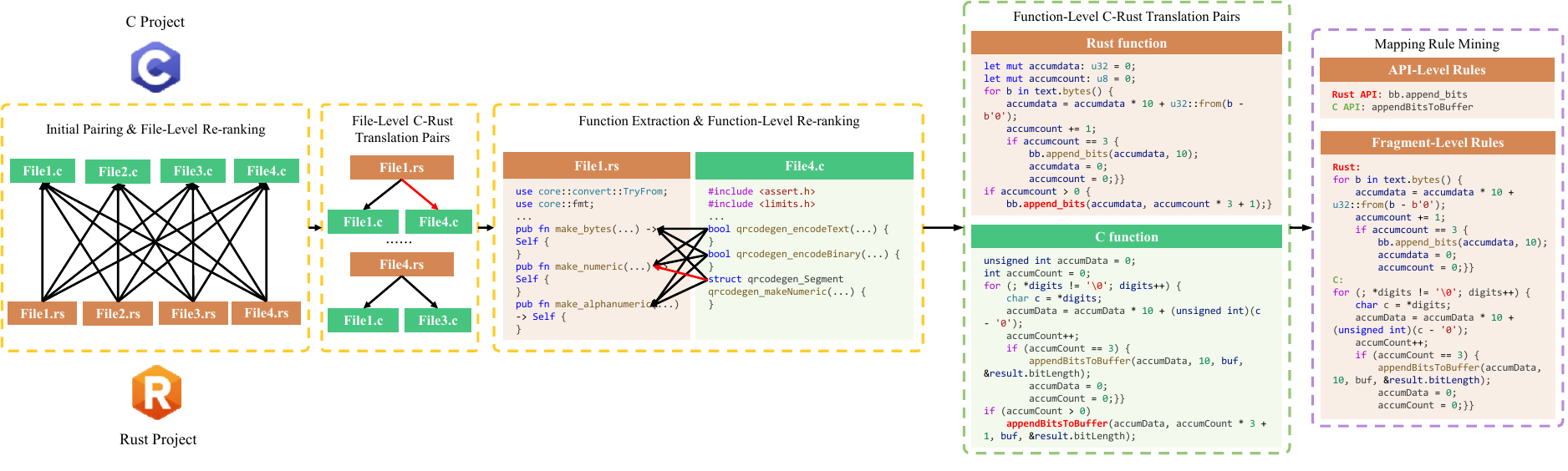}
\caption{Workflow of knowledge-base construction. Candidate C/Rust file pairs are first mined from repository history and code snapshots, then ranked into function-level pairs, and finally summarized into interface-level and fragment-level knowledge.}
\label{fig:kb_construction}
\end{figure*}

\begin{table*}[t]
\centering
\caption{Signals used to mine historical C/Rust migration candidates.}
\label{tab:heuristics}
\scriptsize
\setlength{\tabcolsep}{3pt}
\renewcommand{\arraystretch}{1}
\begin{tabularx}{\textwidth}{@{}>{\centering\arraybackslash}p{0.1\textwidth}>{\centering\arraybackslash}p{0.1\textwidth}>{\raggedright\arraybackslash}X@{}}
\toprule
\textbf{Category} & \textbf{Dimension} & \textbf{Strategy and rationale} \\
\cmidrule(lr){1-3}
\multirow{7}{*}{\makecell[l]{Git history\\(dynamic)}} 
& \multirow{4}{*}{\makecell[l]{Synchronous\\(same commit)}} 
& \textbf{Commit-message keyword matching:} detects explicit migration intent through terms such as \textit{rewrite}, \textit{port}, and \textit{refactor}. \\
& & \textbf{Build-configuration switch:} identifies commits that remove C files and add Rust files within the same build-target definition. \\
& & \textbf{Interface migration:} captures call-site changes where C functions are replaced by Rust-side equivalents in client code. \\
& & \textbf{Code-churn balance:} compares deleted C lines with added Rust lines using commit-level churn statistics to identify likely structural correspondence. \\
\cmidrule(lr){2-3}
& \multirow{3}{*}{\makecell[l]{Asynchronous\\(history window)}} 
& \textbf{Sequential re-implementation:} links a C-file deletion to a later Rust-file creation in the same component within a bounded history window. \\
& & \textbf{Evolutionary coupling:} identifies file pairs that are repeatedly modified together, indicating likely logical dependency. \\
& & \textbf{Developer identity:} prioritizes candidates where the Rust author matches the original C maintainer or a recent contributor. \\
\cmidrule(lr){1-3}
\multirow{3}{*}{\makecell[l]{Codebase\\snapshot\\(static)}} 
& Spatial 
& \textbf{Module colocation:} scans build systems for C and Rust files that coexist in the same module or target. \\
\cmidrule(lr){2-3}
& \multirow{2}{*}{Semantic} 
& \textbf{Key-token overlap:} retains pairs sharing stable identifiers such as constants, error codes, type names, and domain terms. \\
& & \textbf{Shared literals:} matches long string literals, including log messages, error prompts, and protocol strings, that are preserved across languages. \\
\bottomrule
\end{tabularx}
\end{table*}

The mining pipeline proceeds from files to functions. At the file level, candidate C/Rust pairs are collected using the signals in Table~\ref{tab:heuristics}. A lexical retriever first collects likely pairs, and a neural reranking model then orders them by migration relevance~\cite{wang2025jinarerankerv3lateinteractionlistwise}. At the function level, candidate pairs are enumerated within aligned files and ranked again. Retrieved pairs serve as prompt guidance. To reduce direct leakage from evaluation artifacts into retrieval, evaluated target modules are excluded from the historical pairs used to guide their own translations.

Each accepted function-level pair is summarized into two knowledge types. \textbf{Interface-level knowledge} records what Rust-side interface should be reused for a C-side operation, such as a service-manager call, operating-system abstraction wrapper, inter-process communication buffer helper, or domain-specific callback. \textbf{Fragment-level knowledge} records how a recurring local implementation pattern is expressed in Rust, such as error-code propagation, intrusive-list traversal, timeout handling, buffer serialization, or resource release. The distinction lets the prompt separately express what interface to call and how the surrounding logic should be organized.

\subsection{Project Synthesis and Knowledge-Guided Translation}
\label{RustSkeletonConstruction}

Before function bodies are generated, His2Trans constructs a Rust scaffold that defines the migration target. The scaffold specifies the Rust crate layout, module boundaries, recovered type declarations, exported symbols, and placeholder function bodies. This design constrains crate-level invention by giving each later generation step a fixed location, signature, and compilation context.

When build information is available, His2Trans uses it to construct the scaffold from the same source view used by the original project. This reduces inconsistent declarations during scaffold synthesis, while leaving whole-project behavior recovery to later translation and refinement stages.

The scaffold is not intended to be maximally idiomatic Rust at construction time. It makes the migration boundary explicit and carries enough project context for later body generation and refinement. Functions that are called through C/Rust interoperability boundaries may intentionally keep raw pointers or C-compatible layout annotations because changing their signature or layout too early can break mixed-language call chains or test harnesses.

After scaffold construction, His2Trans translates function bodies incrementally. For each target function, the framework assembles local C code, the corresponding Rust signature, visible types, global state, cross-module references, and retrieved historical knowledge. The model then generates a Rust body that is inserted into the scaffold and validated under the current project context.

The translation order follows the scaffold dependency graph. Functions with fewer unresolved dependencies are translated first, so that later functions can refer to already validated definitions. When cycles remain, mutually dependent functions are generated against each other's declared signatures and then checked with whole-crate compiler feedback.

\subsection{Project-Level Agentic Refinement}

The initial translation stage produces a Rust crate, but local body insertion does not guarantee whole-project consistency. Functions may compile individually while still disagreeing about state ownership, resource release order, error boundaries, or unsafe scopes. His2Trans therefore applies project-level refinement after function translation.

The refinement agent operates on the assembled crate with three feedback channels. \textbf{Compilation feedback} exposes unresolved paths, borrow-checker issues, type mismatches, and linking errors. \textbf{Semantic feedback} asks the agent to inspect the assembled artifact against source-level control flow, data flow, error paths, and system interaction protocols. \textbf{Unsafe feedback} identifies unnecessarily broad unsafe scopes and prioritizes rewrites that preserve the migration boundary while moving code toward Rust-native ownership, slice, and indexing patterns. The agent iterates within a fixed budget and may modify multiple functions when a project-level inconsistency cannot be fixed locally.

This stage keeps the migrated crate aligned with the original migration surface. Public interfaces, callbacks, configuration behavior, and system-level protocols remain first-class constraints while the implementation is made more consistent and more Rust-native.

\section{Experimental Setup}
\label{sec:setup}

\subsection{Protocol}

All reported results are computed from Rust artifacts generated by the evaluated method. We do not manually edit translated implementations before measurement. The evaluation harness compiles the generated crate, runs the corresponding tests reused from the original project or its self-contained evaluation harness, and computes compilation, test-pass, and unsafe metrics from the resulting artifact.

Average results are computed at the project level. For each dataset and method, we first compute a percentage for each project and then average those percentages over the fixed project set. This limits the influence of projects with larger test suites or function counts.

Experiments were run on Ubuntu 22.04 with dual Intel Xeon Gold 6430 CPUs, 503 GiB RAM, and four NVIDIA RTX 5880 Ada GPUs. The toolchain used Clang 14.0.0, Rust 1.94.0-nightly, and Python 3.13.5. The offline knowledge extraction stage used \texttt{Qwen2.5-Coder-32B}; the main His2Trans translation and refinement runs used the \texttt{DeepSeek V4 Pro} model archived in the final experiment results.

\subsection{Datasets}

We evaluate on two datasets. We choose OpenHarmony because its GN/Ninja build, HAL layers, and cross-component links stress the recovery of build-time declarations and external symbols beyond isolated algorithmic benchmarks~\cite{OpenHarmony2024}. The \textbf{OpenHarmony module dataset} contains ten modules selected from this build-complex, partially migrated systems ecosystem. These modules cover service management, event dispatch, operating-system abstraction, inter-process communication, shared driver utilities, and audio power-management paths. The \textbf{open-source project dataset} contains eight C projects frequently used in prior C-to-Rust evaluation studies~\cite{Shiraishi2024SmartC2RustIF,dehghan2025translatinglargescalecrepositories,yuan2025projectlevelctorusttranslationsynergistic,nitin2025c2saferrusttransformingcprojects,zhou2025sactorllmdrivencorrectidiomatic}, and is used to evaluate behavior outside the primary ecosystem. Table~\ref{tab:datasets} reports source-level statistics and test counts for the evaluated projects.

\begin{table}[t]
\centering
\caption{Statistics of the datasets used in evaluation.}
\label{tab:datasets}
\footnotesize
\setlength{\tabcolsep}{3pt}
\renewcommand{\arraystretch}{1.06}
\begin{tabularx}{\columnwidth}{@{}>{\centering\arraybackslash}p{0.12\columnwidth}>{\centering\arraybackslash}Xcccc@{}}
\toprule
\textbf{Set} & \textbf{Project} & \#LOC & \#Files & \#Funcs & \#Tests \\
\midrule
\multirow{10}{*}{\rotatebox{90}{OpenHarmony}}
& \texttt{host} & 1,892 & 14 & 127 & 8 \\
& \texttt{appverify\_lite} & 3,544 & 8 & 160 & 8 \\
& \texttt{manager} & 1,535 & 10 & 100 & 4 \\
& \texttt{shared\_541} & 338 & 7 & 26 & 5 \\
& \texttt{posix} & 614 & 6 & 31 & 135 \\
& \texttt{common} & 2,203 & 5 & 84 & 42 \\
& \texttt{core} & 1,899 & 3 & 77 & 30 \\
& \texttt{shared\_12} & 208 & 2 & 7 & 15 \\
& \texttt{osal} & 161 & 1 & 8 & 4 \\
& \texttt{sapm} & 1,439 & 1 & 64 & 8 \\
\midrule
\multirow{8}{*}{\rotatebox{90}{Open-source}}
& \texttt{urlparser} & 427 & 2 & 20 & 3 \\
& \texttt{avl} & 844 & 1 & 29 & 2 \\
& \texttt{ht} & 15 & 1 & 2 & 1 \\
& \texttt{qsort} & 27 & 1 & 3 & 6 \\
& \texttt{buffer} & 353 & 2 & 23 & 14 \\
& \texttt{rgba} & 420 & 2 & 13 & 10 \\
& \texttt{quadtree} & 319 & 5 & 24 & 4 \\
& \texttt{genann} & 459 & 3 & 13 & 12 \\
\bottomrule
\end{tabularx}
\end{table}

\subsection{Baselines}

We compare His2Trans against five baselines. \textbf{C2Rust} represents rule-based transpilation~\cite{Galois2018C2Rust}. \textbf{C2SaferRust} represents an unsafe-reduction pipeline over C-to-Rust output~\cite{nitin2025c2saferrusttransformingcprojects}. \textbf{EvoC2Rust} and \textbf{Tymcrat} represent large-language-model-based project-level migration methods~\cite{wang2026evoc2rustskeletonguidedframeworkprojectlevel,Tymcrat}. For tools whose assumptions do not match OpenHarmony build conventions or scaffold constraints, we run their documented or default project-level workflow under the same harness constraints. We report the resulting failures as observed, without manually adapting their internals. \textbf{Claude Code-only} is a general software-engineering agent baseline. Its prompt provides the original C project, visible tests, an empty Rust output directory, and fairness constraints. It does not require the His2Trans scaffold layout, C-compatible entry points, or historical-knowledge constraints. Claude Code may therefore choose a Rust-native crate layout and public interface. This setting measures the capability and boundary of a general coding agent, not a layout-constrained migration system.

\subsection{Metrics}

We report three metrics. \textbf{Incremental Compilation Pass Rate} is computed by replaying generated function bodies under a common evaluation scaffold. A function is counted as passed if inserting its body preserves compilation under the current project context. \textbf{Test Pass Rate} measures behavior exercised by the available project tests. \textbf{Unsafe Ratio} measures the fraction of Rust source lines that are inside or directly associated with \texttt{unsafe} constructs.

\section{Experimental Evaluation}
\label{sec:evaluation}
    
We evaluate His2Trans through four research questions:

\begin{itemize}[noitemsep, topsep=0pt, leftmargin=1.1em, labelsep=0.5em]
    \item \textbf{RQ1:} How well does His2Trans preserve build-complex migration behavior under OpenHarmony system constraints?
    \item \textbf{RQ2:} Does His2Trans remain effective outside its primary industrial ecosystem?
    \item \textbf{RQ3:} How do the main components affect compilation, testing, and unsafe reduction?
    \item \textbf{RQ4:} How does historical migration knowledge shape translation decisions?
\end{itemize}
  
\subsection{RQ1: OpenHarmony Module Effectiveness}

RQ1 evaluates whether the framework can migrate build-complex systems modules that include C/Rust boundaries, service interfaces, inter-process communication buffers, operating-system abstraction wrappers, and driver-related side effects. We compare His2Trans with C2Rust, C2SaferRust, EvoC2Rust, Tymcrat, and Claude Code-only on the OpenHarmony module dataset.

\subsubsection{Results}

\begin{table*}[t]
\centering
\caption{RQ1 results on OpenHarmony modules. All values are percentages.}
\label{tab:rq1-main-results}
\footnotesize
\setlength{\tabcolsep}{3.5pt}
\renewcommand{\arraystretch}{1.03}
\begin{tabular*}{\textwidth}{@{\extracolsep{\fill}}llrrrrrr@{}}
\toprule
\textbf{Project} & \textbf{Metric} & \textbf{Ours} & \makecell{\textbf{Claude}\\\textbf{Code}} & \textbf{C2Rust} & \makecell{\textbf{C2Safer}\\\textbf{Rust}} & \makecell{\textbf{Evo}\\\textbf{C2Rust}} & \textbf{Tymcrat} \\
\midrule
\multirow{3}{*}{\texttt{host}} & Inc. comp. & 100.00 & 100.00 & 100.00 & 100.00 & 0.00 & 0.00 \\
& Test pass & 100.00 & 75.00 & 75.00 & 75.00 & 0.00 & 0.00 \\
& Unsafe Ratio & 21.72 & 37.84 & 74.72 & 30.08 & 1.24 & 47.12 \\
\addlinespace[0.2em]
\multirow{3}{*}{\texttt{appverify\_lite}} & Inc. comp. & 100.00 & 100.00 & 98.14 & 99.94 & 0.00 & 0.00 \\
& Test pass & 62.50 & 62.50 & 0.00 & 0.00 & 0.00 & 0.00 \\
& Unsafe Ratio & 16.33 & 6.68 & 52.45 & 7.24 & 0.00 & 7.29 \\
\addlinespace[0.2em]
\multirow{3}{*}{\texttt{manager}} & Inc. comp. & 100.00 & 100.00 & 100.00 & 0.00 & 0.00 & 0.00 \\
& Test pass & 100.00 & 0.00 & 0.00 & 0.00 & 0.00 & 0.00 \\
& Unsafe Ratio & 23.74 & 34.04 & 50.20 & 36.40 & 1.38 & 43.33 \\
\addlinespace[0.2em]
\multirow{3}{*}{\texttt{shared\_541}} & Inc. comp. & 100.00 & 100.00 & 100.00 & 100.00 & 0.00 & 0.00 \\
& Test pass & 100.00 & 100.00 & 100.00 & 60.00 & 0.00 & 0.00 \\
& Unsafe Ratio & 12.66 & 31.25 & 32.15 & 17.53 & 1.41 & 30.17 \\
\addlinespace[0.2em]
\multirow{3}{*}{\texttt{posix}} & Inc. comp. & 100.00 & 100.00 & 100.00 & 100.00 & 0.00 & 0.00 \\
& Test pass & 100.00 & 100.00 & 100.00 & 0.00 & 0.00 & 0.00 \\
& Unsafe Ratio & 26.27 & 38.73 & 54.05 & 1.01 & 1.41 & 48.92 \\
\addlinespace[0.2em]
\multirow{3}{*}{\texttt{common}} & Inc. comp. & 100.00 & 100.00 & 97.70 & 100.00 & 0.00 & 0.00 \\
& Test pass & 100.00 & 100.00 & 0.00 & 0.00 & 0.00 & 0.00 \\
& Unsafe Ratio & 14.09 & 35.96 & 53.53 & 37.73 & 0.90 & 47.39 \\
\addlinespace[0.2em]
\multirow{3}{*}{\texttt{core}} & Inc. comp. & 100.00 & 100.00 & 100.00 & 0.00 & 0.00 & 0.00 \\
& Test pass & 86.67 & 83.33 & 100.00 & 0.00 & 0.00 & 0.00 \\
& Unsafe Ratio & 18.41 & 39.31 & 62.40 & 45.43 & 0.95 & 43.20 \\
\addlinespace[0.2em]
\multirow{3}{*}{\texttt{shared\_12}} & Inc. comp. & 100.00 & 100.00 & 100.00 & 100.00 & 0.00 & 0.00 \\
& Test pass & 100.00 & 80.00 & 100.00 & 86.67 & 0.00 & 0.00 \\
& Unsafe Ratio & 7.93 & 40.56 & 94.45 & 30.76 & 1.60 & 30.84 \\
\addlinespace[0.2em]
\multirow{3}{*}{\texttt{osal}} & Inc. comp. & 100.00 & 100.00 & 100.00 & 100.00 & 0.00 & 90.91 \\
& Test pass & 100.00 & 100.00 & 100.00 & 100.00 & 0.00 & 0.00 \\
& Unsafe Ratio & 12.85 & 32.70 & 42.04 & 28.04 & 1.60 & 34.73 \\
\addlinespace[0.2em]
\multirow{3}{*}{\texttt{sapm}} & Inc. comp. & 100.00 & 100.00 & 100.00 & 100.00 & 0.00 & 0.00 \\
& Test pass & 100.00 & 100.00 & 100.00 & 87.50 & 0.00 & 0.00 \\
& Unsafe Ratio & 9.54 & 39.66 & 65.07 & 42.04 & 1.07 & 47.26 \\
\midrule
\textbf{Average} & Inc. comp. & \textbf{100.00} & \textbf{100.00} & \underline{99.58} & 79.99 & 0.00 & 9.09 \\
& Test pass & \textbf{94.92} & \underline{80.08} & 67.50 & 40.92 & 0.00 & 0.00 \\
& Unsafe Ratio & \underline{16.35} & 33.67 & 58.11 & 27.63 & \textbf{1.16} & 38.03 \\
\bottomrule
\end{tabular*}
\vspace{1pt}
\begin{minipage}{\textwidth}
\footnotesize Inc. comp. denotes incremental compilation. Higher is better for incremental compilation and test pass rate; lower is better for Unsafe Ratio. In the average rows, bold and underline denote the best and second-best results, respectively.
\end{minipage}
\end{table*}

Table~\ref{tab:rq1-main-results} reports the OpenHarmony results. In this dataset, His2Trans is the only evaluated method that combines full incremental compilation, a high test pass rate, and a low Unsafe Ratio. It reaches 100.00\% incremental compilation, a 94.92\% test pass rate, and a 16.35\% Unsafe Ratio. Claude Code also reaches 100.00\% incremental compilation, but its test pass rate is 80.08\% and its Unsafe Ratio is 33.67\%. C2Rust reaches 99.58\% incremental compilation, 67.50\% test pass rate, and 58.11\% Unsafe Ratio. C2SaferRust reduces unsafe relative to C2Rust, but its test pass rate remains 40.92\%.

\subsubsection{Analysis}

The OpenHarmony results indicate that project-level systems migration needs explicit constraints beyond compilation, because the tests exercise system-facing contracts such as event delivery, service-manager access, operating-system abstraction return codes, inter-process communication field order, and driver-side state updates.

The observed baseline errors are consistent with this boundary-mismatch interpretation. C2Rust nearly preserves compilability because it lowers C control flow, data layout, and pointer operations into Rust with high structural correspondence. However, the OpenHarmony tests also exercise project-observable contracts that are not checked by Rust compilation alone, including initialization order, callback registration, buffer-field order, domain error codes, and driver-facing state. A mechanically translated artifact can therefore compile while remaining misaligned with these surrounding protocols, which explains the gap between C2Rust's 99.58\% incremental compilation rate and 67.50\% test pass rate. C2SaferRust reduces some unsafe patterns after mechanical translation, but the same protocol-level errors can remain or become more fragile when low-level boundary code is rewritten without project-specific migration knowledge.

EvoC2Rust and Tymcrat exhibit a different error pattern in this dataset. EvoC2Rust's OpenHarmony output is not empty: for example, the \texttt{host} run contains a \texttt{Cargo.toml}, translated \texttt{src/*.rs} files, and hundreds of Rust function bodies. The incremental-compilation checker can scan these functions, but it cannot validate any restored body because neither the generated crate nor the function-stubbed replay context compiles. The observed errors include untranslated C macro-loop syntax, Rust keyword-derived fields, duplicate macro definitions, missing OpenHarmony interface types, incompatible string macro calls, and invalid static initialization. Thus the 0.00\% incremental compilation rate is caused by the absence of a compilable Rust project context for per-function replay. The 0.00\% Test Pass Rate follows from the same noncompiling artifact before static-library export and test linking. Tymcrat has a smaller partial-build footprint, but it still does not reach a linked and executable OpenHarmony test artifact.

The comparison with Claude Code separates compilation from migration fidelity. Claude Code obtains 100.00\% incremental compilation, but its lower test pass rate and higher Unsafe Ratio show that compilation alone is not a sufficient objective for systems migration. The archived OpenHarmony failures include cases where the generated crate internalizes C service-manager entry points into a Rust object model, rewrites the dynamic-loader protocol into a Rust-native library lookup, or replaces external serialization APIs with standalone buffers. These choices can be locally plausible, but they weaken system-observable migration boundaries that are not enforced by the Rust compiler. His2Trans is designed around this stricter objective.

\begin{figure}[t]
\centering
\includegraphics[width=0.9\linewidth, trim={0pt 329pt 746pt 0pt}, clip]{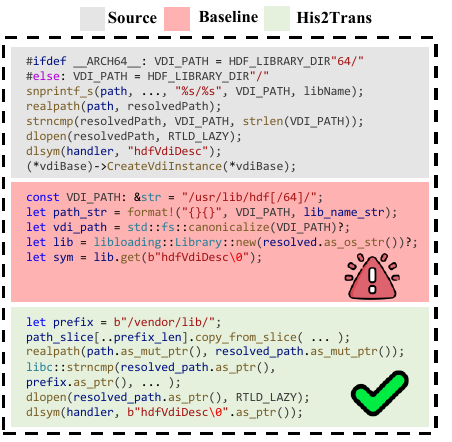}
\caption{Representative OpenHarmony VDI boundary case in \texttt{host}. Claude Code-only uses a conventional \texttt{/usr/lib/hdf[/64]/} lookup, whereas His2Trans preserves the OpenHarmony deployment path and native symbol-resolution protocol. The archived \texttt{host} tests are 6/8 for Claude Code-only and 8/8 for His2Trans.}
\label{fig:rq1-vdi-boundary}
\end{figure}

Figure~\ref{fig:rq1-vdi-boundary} gives a concrete example from \texttt{host}. The source VDI loader constrains library lookup to the OpenHarmony HDF deployment path, canonicalizes the path before loading, resolves the exported \texttt{hdfVdiDesc} symbol, and initializes the VDI instance through the native ABI. Claude Code-only produces a plausible Rust loader, but uses a conventional \texttt{/usr/lib/hdf/} lookup boundary, which is consistent with the failed sample VDI loading tests. His2Trans keeps the deployment path boundary and symbol-resolution protocol, matching the observed \texttt{host} outcome of 8/8 tests compared with 6/8 for Claude Code-only.

\subsubsection{Answer to RQ1}

On the OpenHarmony module dataset, His2Trans achieves the highest Test Pass Rate and a lower Unsafe Ratio than Claude Code, C2Rust, C2SaferRust, and Tymcrat. The results support historical knowledge reuse and project-level agentic refinement for build-complex systems modules.

\subsection{RQ2: Open-Source Project Effectiveness}

RQ2 evaluates whether His2Trans remains useful beyond its primary industrial-context setting, using the open-source project dataset described in Section~\ref{sec:setup}.

\subsubsection{Results}

\begin{table*}[t]
\centering
\caption{RQ2 results on open-source projects. All values are percentages.}
\label{tab:rq2-main-results}
\footnotesize
\setlength{\tabcolsep}{3.5pt}
\renewcommand{\arraystretch}{1.03}
\begin{tabular*}{\textwidth}{@{\extracolsep{\fill}}llrrrrrr@{}}
\toprule
\textbf{Project} & \textbf{Metric} & \textbf{Ours} & \makecell{\textbf{Claude}\\\textbf{Code}} & \textbf{C2Rust} & \makecell{\textbf{C2Safer}\\\textbf{Rust}} & \makecell{\textbf{Evo}\\\textbf{C2Rust}} & \textbf{Tymcrat} \\
\midrule
\multirow{3}{*}{\texttt{urlparser}} & Inc. comp. & 100.00 & 100.00 & 100.00 & 100.00 & 30.43 & 0.00 \\
& Test pass & 100.00 & 100.00 & 100.00 & 0.00 & 0.00 & 0.00 \\
& Unsafe Ratio & 10.57 & 0.00 & 65.11 & 49.73 & 1.46 & 0.71 \\
\addlinespace[0.2em]
\multirow{3}{*}{\texttt{avl}} & Inc. comp. & 100.00 & 100.00 & 100.00 & 100.00 & 48.00 & 0.00 \\
& Test pass & 100.00 & 100.00 & 100.00 & 0.00 & 0.00 & 0.00 \\
& Unsafe Ratio & 7.67 & 32.32 & 55.55 & 36.80 & 1.47 & 65.92 \\
\addlinespace[0.2em]
\multirow{3}{*}{\texttt{ht}} & Inc. comp. & 100.00 & 100.00 & 100.00 & 100.00 & 0.00 & 97.22 \\
& Test pass & 100.00 & 100.00 & 100.00 & 0.00 & 0.00 & 0.00 \\
& Unsafe Ratio & 1.15 & 0.00 & 34.18 & 3.70 & 0.00 & 1.98 \\
\addlinespace[0.2em]
\multirow{3}{*}{\texttt{qsort}} & Inc. comp. & 100.00 & 100.00 & 100.00 & 100.00 & 100.00 & 100.00 \\
& Test pass & 100.00 & 100.00 & 100.00 & 100.00 & 100.00 & 0.00 \\
& Unsafe Ratio & 7.03 & 0.00 & 22.67 & 0.00 & 1.68 & 69.23 \\
\addlinespace[0.2em]
\multirow{3}{*}{\texttt{buffer}} & Inc. comp. & 100.00 & 100.00 & 100.00 & 100.00 & 0.00 & 95.92 \\
& Test pass & 100.00 & 100.00 & 100.00 & 100.00 & 0.00 & 0.00 \\
& Unsafe Ratio & 15.83 & 0.00 & 50.09 & 19.11 & 1.46 & 0.27 \\
\addlinespace[0.2em]
\multirow{3}{*}{\texttt{rgba}} & Inc. comp. & 100.00 & 100.00 & 100.00 & 100.00 & 0.00 & 100.00 \\
& Test pass & 100.00 & 100.00 & 100.00 & 0.00 & 0.00 & 0.00 \\
& Unsafe Ratio & 3.84 & 0.00 & 0.39 & 0.00 & 1.52 & 0.00 \\
\addlinespace[0.2em]
\multirow{3}{*}{\texttt{quadtree}} & Inc. comp. & 100.00 & 100.00 & 100.00 & 100.00 & 0.00 & 0.00 \\
& Test pass & 100.00 & 100.00 & 100.00 & 75.00 & 0.00 & 0.00 \\
& Unsafe Ratio & 10.68 & 0.00 & 47.65 & 44.38 & 1.56 & 0.00 \\
\addlinespace[0.2em]
\multirow{3}{*}{\texttt{genann}} & Inc. comp. & 100.00 & 100.00 & 100.00 & 100.00 & 0.00 & 0.00 \\
& Test pass & 100.00 & 100.00 & 100.00 & 100.00 & 0.00 & 0.00 \\
& Unsafe Ratio & 11.99 & 0.00 & 67.44 & 39.81 & 1.50 & 0.23 \\
\midrule
\textbf{Average} & Inc. comp. & \textbf{100.00} & \textbf{100.00} & \textbf{100.00} & \textbf{100.00} & 22.30 & \underline{49.14} \\
& Test pass & \textbf{100.00} & \textbf{100.00} & \textbf{100.00} & \underline{46.88} & 12.50 & 0.00 \\
& Unsafe Ratio & 8.59 & \underline{4.04} & 42.88 & 24.19 & \textbf{1.33} & 17.29 \\
\bottomrule
\end{tabular*}
\vspace{1pt}
\begin{minipage}{\textwidth}
\footnotesize Inc. comp. denotes incremental compilation. Higher is better for incremental compilation and test pass rate; lower is better for Unsafe Ratio. In the average rows, bold and underline denote the best and second-best results, respectively.
\end{minipage}
\end{table*}

Table~\ref{tab:rq2-main-results} reports that His2Trans, Claude Code, and C2Rust all achieve 100.00\% incremental compilation and 100.00\% Test Pass Rate on the open-source project dataset. In this dataset, the observable test surface is compact enough for multiple methods to recover the tested behavior.

Unsafe Ratio separates the methods more clearly. C2Rust's ratio is 42.88\%, reflecting the unsafe burden of mechanically preserving C structure even when tests pass. His2Trans reduces the ratio to 8.59\%. Among the methods that also reach 100.00\% incremental compilation and 100.00\% test pass rate, Claude Code achieves the lowest Unsafe Ratio, 4.04\%. This is consistent with the baseline setting: Claude Code may choose Rust-native interfaces and crate boundaries rather than preserving the original C migration surface.

\subsubsection{Analysis}

The successful RQ2 rows reflect different migration styles. C2Rust preserves behavior through extensive unsafe code. Claude Code often produces a cleaner Rust-native library because it is free to reorganize interfaces and crate boundaries. His2Trans occupies a different point in the design space: it retains source-level behavior at constrained interfaces while using Rust-native ownership and slices where the migration permits, reducing Unsafe Ratio relative to C2Rust.

The error profile is correspondingly different from RQ1. The open-source projects are smaller and more self-contained, so direct structural preservation is often sufficient for the tested behavior; this is why C2Rust reaches full incremental compilation and full test pass rate despite its high unsafe ratio. For EvoC2Rust, Tymcrat, and C2SaferRust, the weaker rows come from concrete migration errors in the translated artifacts, including incomplete project reconstruction, missing exported APIs, and post-translation safety rewrites that break required entry points or behavior. In this setting, His2Trans benefits less from industrial system knowledge than in RQ1, but still retains source-level behavior where interfaces constrain it while reducing unsafe code relative to C2Rust.

The same pattern appears in a concrete RQ2 case with runtime consequences. In \texttt{ht}, the C source defines a private \texttt{hash\_key(const char *key)} helper whose loop stops at the first \texttt{NUL} byte. The available test harness cannot directly call this private helper, so it does not constrain the behavior on strings that contain an embedded \texttt{NUL}. Claude Code-only translates the helper as \texttt{hash\_key(data: \&[u8])} and intentionally hashes every byte in the slice; its own test asserts that \texttt{hash\_key(b"ab\textbackslash{}0cd")} differs from \texttt{hash\_key(b"ab")}. His2Trans keeps the C-string interpretation by translating the helper over \texttt{CStr} and iterating over \texttt{to\_bytes()}, which hashes only the bytes before the first terminator. For the same underlying memory \texttt{ab\textbackslash{}0cd}, the C source and His2Trans therefore hash \texttt{ab}, whereas the Claude Code-only artifact hashes the full byte slice. This case shows a source-level runtime detail that compact tests can miss.

\begin{figure}[t]
\centering
\includegraphics[width=0.9\linewidth, trim={0pt 235pt 746pt 0pt}, clip]{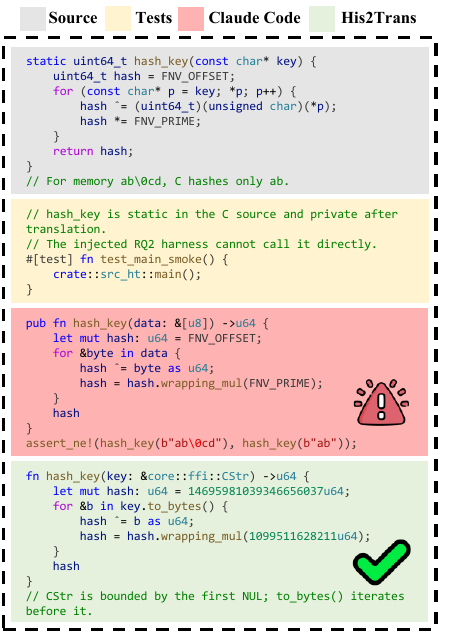}
\caption{Representative untested runtime-semantics case in \texttt{ht}. Claude Code-only rewrites the private C-string hash as a byte-slice hash and includes bytes after an embedded \texttt{NUL}; His2Trans keeps the C-string interpretation through \texttt{CStr} and \texttt{to\_bytes()}.}
\label{fig:rq2-ht-hash}
\end{figure}

\subsubsection{Answer to RQ2}

His2Trans achieves 100.00\% incremental compilation and 100.00\% Test Pass Rate on the open-source project dataset. It reduces Unsafe Ratio relative to C2Rust while retaining source-level behavior at constrained interfaces; the \texttt{ht} case shows how that behavior can include runtime details that compact tests do not exercise.

\subsection{RQ3: Ablation Study}

\begin{table}[t]
\centering
\caption{Ablation results on the OpenHarmony module dataset.}
\label{tab:rq3-merged}
\footnotesize
\setlength{\tabcolsep}{3pt}
\renewcommand{\arraystretch}{1.0}
\begin{tabularx}{\linewidth}{@{}>{\centering\arraybackslash}Xccc@{}}
\toprule
\textbf{Setting} & \makecell{\textbf{Incremental}\\\textbf{compilation}} & \makecell{\textbf{Test}\\\textbf{Pass Rate}} & \makecell{\textbf{Unsafe}\\\textbf{Ratio}} \\
\midrule
Knowledge-guided function translation & 95.82\% & 39.29\% & 15.95\% \\
With compiler-driven per-function repair & 100.00\% & 94.92\% & 22.52\% \\
With project-level agentic refinement & 100.00\% & 94.92\% & 16.35\% \\
\bottomrule
\end{tabularx}
\end{table}

RQ3 studies how the main His2Trans stages affect results on the OpenHarmony module dataset. The first setting uses historical-knowledge-guided function translation without compiler-driven repair. The second setting adds compiler-driven per-function repair. The third setting adds project-level agentic refinement and corresponds to the full framework.

\subsubsection{Results}

Initial function translation reaches 95.82\% incremental compilation, but its test pass rate is only 39.29\%. Most functions can be inserted into the scaffold, yet many generated bodies do not preserve behavior sufficiently to pass project tests. Compiler-driven per-function repair is associated with the main recovery in behavior: it raises incremental compilation to 100.00\% and test pass rate to 94.92\%.

The cost of local repair is visible in unsafe code. Adding compiler-driven per-function repair increases Unsafe Ratio from 15.95\% to 22.52\%, indicating that local repair often chooses conservative low-level fixes to satisfy compilation and tests. The final refinement stage does not further raise the aggregate Test Pass Rate, which remains 94.92\% rather than 100.00\%. This unchanged score should be read narrowly: Test Pass Rate did not capture all source-level semantic improvements made by project-level refinement. The full framework reduces Unsafe Ratio to 16.35\% without lowering Test Pass Rate, while replacing conservative low-level repairs where project-level feedback supports safer ownership, slice, or indexing patterns.

\subsubsection{Analysis}

\begin{figure}[t]
\centering
\includegraphics[width=0.9\linewidth, trim={0pt 327pt 746pt 0pt}, clip]{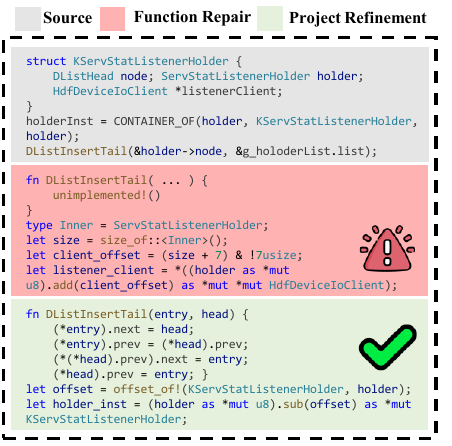}
\caption{Representative project-level refinement case in \texttt{manager}. Function repair compiles with incomplete DList handling and an offset guess; project refinement restores \texttt{offset\_of}-based layout recovery and concrete DList updates without changing the aggregate Test Pass Rate.}
\label{fig:rq3-listener-holder}
\end{figure}

Figure~\ref{fig:rq3-listener-holder} shows this effect in \texttt{manager}. The source listener-holder implementation embeds a public \texttt{ServStatListenerHolder} inside \texttt{KServStatListenerHolder}, recovers the outer object with a container-of layout relation, and maintains the holders in a global DList. After per-function repair, the function-repair version still compiles but leaves DList helpers as \texttt{unimplemented!()} and recovers the listener client through an alignment-based offset guess from the inner holder. Project-level refinement replaces this brittle local repair with an explicit \texttt{offset\_of} layout recovery and concrete DList pointer updates. The \texttt{manager} tests remain 4/4 in both the function-repair and full-framework artifacts, so the aggregate Test Pass Rate does not move, but the final artifact is closer to the source-level structure.

\subsubsection{Answer to RQ3}

Compiler-driven repair is the key factor associated with recovering behavior exercised by the tests. Project-level refinement then improves source-level semantic consistency and reduces unsafe code without lowering the Test Pass Rate; the unchanged aggregate Test Pass Rate does not fully reflect these semantic improvements.

\subsection{RQ4: Historical Knowledge Reuse Analysis}

RQ4 qualitatively examines how historical migration knowledge shapes translation decisions. We focus on archived OpenHarmony translations whose correctness depends on system-facing protocols. The cases follow a concise evidence chain: knowledge, source, baseline, and His2Trans.

\subsubsection{Results}

\begin{figure}[t]
\centering
\includegraphics[width=0.9\linewidth, trim={0pt 289pt 746pt 0pt}, clip]{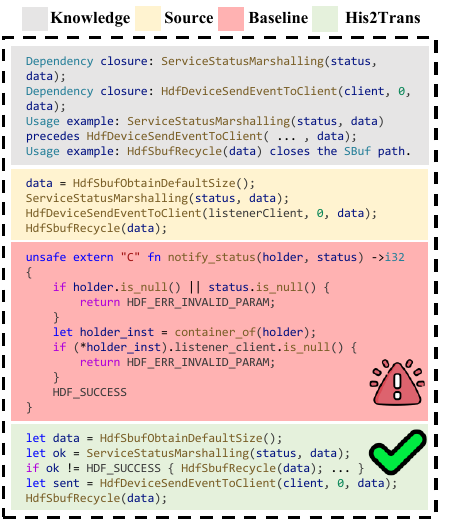}
\caption{Historical-knowledge reuse case in \texttt{manager}. Retrieved event-notification knowledge marks SBuf marshalling and \texttt{HdfDeviceSendEventToClient} as a system-visible path. Claude Code-only reduces the callback to local success, whereas His2Trans keeps construction, delivery, and cleanup.}
\label{fig:rq4-service-event}
\end{figure}

Figure~\ref{fig:rq4-service-event} shows a service-status notification case from \texttt{manager}. The knowledge supplied to the translation prompt highlights the relevant dependency and usage context: a status update is marshalled into an HDF SBuf and then delivered through \texttt{HdfDeviceSendEventToClient}. The target source follows this protocol by allocating a default SBuf, calling \texttt{ServiceStatusMarshalling}, sending the event to the registered client, and recycling the buffer. The retrieved knowledge gives the translator a reusable constraint: event notification must preserve an externally visible data path.

\begin{figure}[!h]
\centering
\includegraphics[width=0.9\linewidth, trim={0pt 256pt 746pt 0pt}, clip]{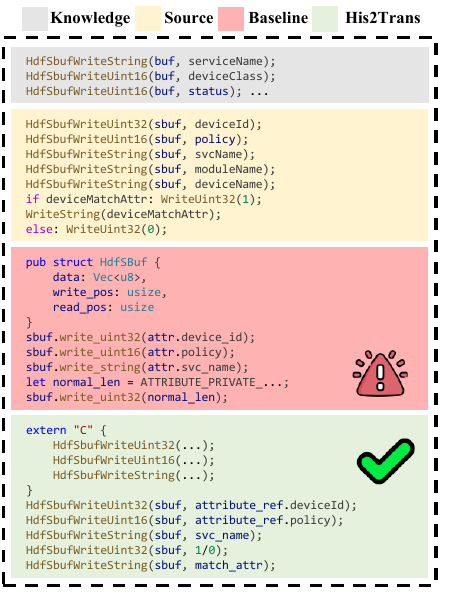}
\caption{Historical-knowledge reuse case in \texttt{shared\_12}. Retrieved SBuf-marshalling context identifies ordered HDF SBuf writes as an external wire-format boundary. Claude Code-only uses a standalone Rust byte buffer and passes 12/15 tests, whereas His2Trans keeps the external SBuf calls and passes 15/15 tests.}
\label{fig:rq4-sbuf-wire}
\end{figure}

Figure~\ref{fig:rq4-sbuf-wire} shows a second case from \texttt{shared\_12}, where the target function serializes device attributes into an HDF SBuf. The prompt context contains SBuf-marshalling usage evidence, and the target source fixes the field order as \texttt{deviceId}, \texttt{policy}, \texttt{svcName}, \texttt{moduleName}, \texttt{deviceName}, and the optional \texttt{deviceMatchAttr} payload. This context guides the translation toward treating SBuf as a project wire-format boundary rather than as a crate-local byte-buffer abstraction.

\subsubsection{Analysis}

The generated artifacts expose the difference. In \texttt{manager}, Claude Code-only keeps the callback type and local null checks but reduces the body to returning \texttt{HDF\_SUCCESS}. His2Trans instead keeps the system-visible event path and its cleanup logic.

In \texttt{shared\_12}, Claude Code-only implements \texttt{HdfSBuf} as a crate-local \texttt{Vec<u8>} buffer and passes 12/15 tests, with failures in \texttt{DevAttributeSerializeTest004}, \texttt{DevAttributeSerializeTest005}, and \texttt{DevAttributeSerializeTest014}. His2Trans keeps the external \texttt{HdfSbufWrite*} and \texttt{HdfSbufRead*} calls, preserves the string allocation/release path, and passes 15/15 tests.

\subsubsection{Answer to RQ4}

Historical migration knowledge shapes translation decisions by turning prior accepted C/Rust migration patterns and usage context into reusable interface constraints. In the archived cases, His2Trans uses this evidence to preserve both an HDF event-notification boundary and an HDF SBuf wire-format boundary.

\subsection{Overall Discussion}

Across the OpenHarmony and open-source datasets, the evidence for His2Trans does not rest on any single metric in isolation. The framework maintains compilation feasibility, preserves observable behavior, and reduces measured unsafe code while keeping an explicit migration boundary. This combination is important for gradual C/Rust migration, where a translated Rust artifact must remain interoperable with existing code and future maintenance expectations.

\section{Threats to Validity}
\label{sec:threats}
 
\noindent\textbf{Test adequacy.}
Test Pass Rate measures behavior exercised by the available tests. The OpenHarmony and open-source project suites cover important behavior, but they cannot exhaustively cover concurrency interleavings, hardware states, exceptional paths, undefined behavior in the original C code, or all deployment-specific I/O modes.

\noindent\textbf{Average reporting.}
We report project-level averages so that each project contributes equally. This limits domination by projects with larger test suites or more functions.

\noindent\textbf{Potential data leakage.}
Although evaluated target modules are excluded from retrieval, indirect leakage cannot be fully ruled out. Public mirrors, upstream history, similar modules, or large language model pretraining data may contain overlapping implementation patterns.

\section{Related Work}
\label{sec:related_work}

\noindent\textbf{Rule-based C-to-Rust translation.}
Rule-based systems translate C through compiler front-ends, abstract-syntax-tree rewriting, and static analyses. C2Rust~\cite{Galois2018C2Rust} is the most widely used foundation, while follow-up work studies how to reduce unsafety through ownership analysis, aliasing constraints, tagged-union translation, and specialized rewrites~\cite{10.1145/3485498,10.1007/978-3-031-37709-9_22,emre2023aliasing,ling2022rust,hong2024tag,hong2025forcrat}. These methods are attractive because they preserve C structure, but the resulting Rust often remains pointer-heavy and difficult to maintain.

\noindent\textbf{Large-language-model-based translation and repair.}
Large-language-model-based C-to-Rust methods use model generation to produce more idiomatic Rust and to repair errors using feedback from tests, compilation, specifications, or static analyses~\cite{eniser2025translatingrealworldcodellms,yang2024exploring,nitin2025spectraenhancingcodetranslation,xu2025optimizing,luo2025integratingrulessemanticsllmbased,farrukh2025safetransllmassistedtranspilationc,11126570,yang2024vertverifiedequivalentrust}. These approaches show that language models can recover semantics that are difficult for rule-based rewriting alone. Their limitation is that local generation and feedback are not enough when a project contains build-specific declarations, C/Rust interoperability entries, or long-lived ecosystem conventions.

\noindent\textbf{Project-level migration.}
Recent work has moved toward project-level migration with skeletons, dependency graphs, knowledge graphs, and iterative validation. EvoC2Rust uses a skeleton-guided framework for project-level translation~\cite{wang2026evoc2rustskeletonguidedframeworkprojectlevel}; PTRMAPPER integrates knowledge graphs and large language models~\cite{yuan2025projectlevelctorusttranslationsynergistic}; RustMap combines dependency guidance and LLM-based translation~\cite{10.1007/978-3-032-00828-2_16}; LLMigrate and Tymcrat further explore project-level and type-migration settings~\cite{liu2025llmigratetransforminglazylarge,Tymcrat}. His2Trans follows this project-level direction, but uses historical C/Rust migration pairs as reusable guidance during translation and refinement.

\noindent\textbf{Agentic C-to-Rust migration.}
A newer line of preprints and submissions treats C-to-Rust migration as an agentic workflow, where tools plan, edit, validate, and repair code across multiple steps rather than issuing a single translation call. ORBIT uses dependency-guided orchestration and specialized agents for autonomous project-level transpilation~\cite{farrukh2026orbit}. RustPrint uses architecture-aware documentation as a migration blueprint and repair signal~\cite{leanh2026rustprint}. ENCRUST combines encapsulated substitution with agentic refinement on a live scaffold~\cite{sim2026encrust}. ACToR uses a translator--discriminator loop to find divergent inputs and refine translations~\cite{li2025actor}. LAC2R explores search-based multi-trajectory refinement for safe C-to-Rust translation~\cite{sim2025lac2r}. Rustify similarly explored workflow-guided multi-agent repository translation~\cite{wang2025rustifywithdrawn}.

\noindent\textbf{Rust adoption and mixed-language systems.}
The adoption of Rust in Linux, Android, ChromeOS, and Firefox illustrates that practical migration is gradual and mixed-language rather than a single complete rewrite~\cite{Ojeda2022RustLinux,AndroidMemorySafetyAOSP,ChromiumOSDevelopmentBasics,LinuxKernelRustQuickStart,FirefoxRustCppInterop}. Such environments motivate preserving C/Rust interoperability boundaries, inter-process communication formats, and service boundaries during migration. His2Trans targets this setting by using historical C/Rust pairs as reusable evidence of how previous migrations handled local interfaces and system protocols.

\section{Conclusion}
\label{sec:conclusion}

This paper presented His2Trans, a knowledge-guided agentic framework for project-level C-to-Rust migration. His2Trans combines historical interface-level and fragment-level migration knowledge with project-level agentic refinement. On ten OpenHarmony modules, it achieves 100.00\% incremental compilation, a 94.92\% Test Pass Rate, and a 16.35\% Unsafe Ratio. On eight open-source C projects, it achieves 100.00\% incremental compilation and 100.00\% Test Pass Rate while reducing Unsafe Ratio relative to C2Rust. The ablation study indicates that local compiler feedback recovers behavior measured by the tests, while project-level refinement improves source-level semantic consistency and reduces measured unsafe code. The Claude Code-only analysis further shows why migration-specific constraints matter: a general coding agent can produce compiling Rust crates, but OpenHarmony tests expose weaknesses in system-specific protocol preservation. In the evaluated setting, historical migration knowledge and project-level agentic refinement provide practical mechanisms for producing compilable and testable Rust artifacts with reduced measured unsafe code.

\section*{Data Availability Statement}
The raw data, test code, and framework source code used in this study are available at: \url{https://anonymous.4open.science/r/His2Trans-F1F0/}.

\bibliographystyle{style/IEEEtran}
\bibliography{refs/IEEEabrv,refs/main}

\end{document}